\documentclass[
    ,final            
  ]
  {aipproc}

\layoutstyle{8x11double}

\begin{document}

\title{Measurements of Black Hole Spins and Tests of Strong-Field General
Relativity}

\author{Dimitrios Psaltis}{
  address={Physics Department, University of Arizona, 1118 E.
 4th St., Tucson, AZ 85721}
}

\begin{abstract}
Fast variability studies of accreting black holes in the Galaxy offer
us a unique opportunity to measure the spins of black holes and test
the strong-field behavior of general relativity. In this review, I 
summarize the arguments often used in attempts of measuring the spins of
black holes, concentrating on their theoretical foundations. I
also argue that X-ray studies of accreting black holes will be able to
provide in the future strong constraints on deviations from general
relativity in the strong-field regime.
\end{abstract}

\maketitle


\section{Introduction}

Astrophysical black holes in general relativity are characterized by
two quantities, their masses and spins, which determine uniquely the
properties of their gravitational fields. As a result, both can, in
principle, be measured by experiments involving test-particle orbits
in their exterior spacetimes.

This approach has been very successful in measuring black-hole masses
both in galactic systems (McClintock \& Remillard 2003) and in the
centers of galaxies (Sch{\" o}del et al.\ 2002). However, the imprint
of the spins of black holes on their spacetimes, i.e., the dragging of
inertial frames, is very weak at the large distances from the
horizons, where the observed orbits reside. It is expected that the
detection of gravitational waves from close, inspiraling compact
objects with LIGO and LISA will allow for a complete mapping of the
spacetimes of the objects and hence for the measurement of black-hole
spins (see, e.g., Hughes 2003). However, only a small fraction of
mostly supermassive black holes exist in the near universe in
configurations that will allow such studies. 

Most of the black holes we observe today are visible because they
accrete matter from their companions or the surrounding medium. The
intense X-ray radiation we detect is generated in a region only a few
Schwarzschild radii around the black-hole horizons. As a result, these
X-ray photons carry with them the signatures of the strong
gravitational fields in which they are produced and hence information
regarding the masses and spins of the black holes.

In this review, I discuss the potential of measuring black-hole spins
and confirming the predictions of general relativity, using their
rapid-variability properties. In particular, I concentrate on the
various methods of inferring black-hole spins that are based on the
observations of constant-frequency quasi-periodic oscillations from
galactic black-hole binaries (QPOs; see Remillard, this volume). The
frequencies of these QPOs depend very weakly on the observed X-ray
flux and, for this reason, it is believed that they are determined
mostly by gravity and not by the hydrodynamic properties of the
accretion flows, such as their temperatures and densities.

\section{QPOs and The Spins of Black Holes}

A number of different arguments have been used recently in the
literature for inferring the spins of black holes from the frequencies
of observed QPOs. In this section, I describe the theoretical foundations
of three types of arguments that appear to depend the least on the
specifics of theoretical models. 

\subsection{The maximum QPO frequency}

In general relativity, the orbits of test particles around a spinning
black hole are characterized by three frequencies: the azimuthal
(Keplerian) frequency $f_\phi$, which is simply equal to the inverse
of the orbital period; the (radial) epicyclic frequency $\kappa$,
which for nearly-circular orbits is equal to the frequency of radial
oscillations around the mean orbit; and the (vertical) epicyclic
frequency $f_\perp$. All three frequencies depend on the mass and spin
of the black hole, as well as on the radius of the orbit.

\begin{figure}
  \includegraphics[height=.3\textheight]{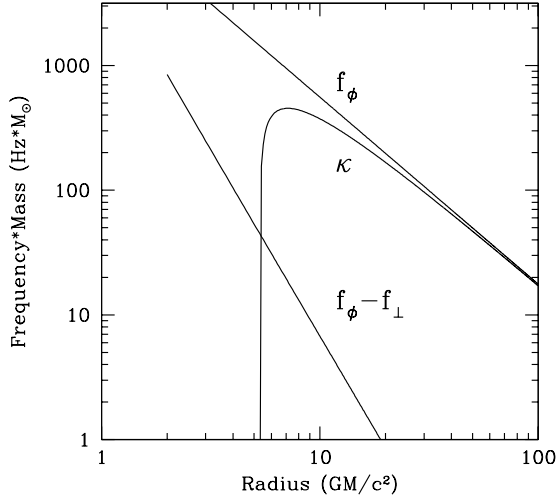} 
  \caption{The azimuthal ($f_\phi$), radial epicyclic ($\kappa$), 
and vertical ($f_\perp$) frequencies at different radii around a
black hole with spin parameter $a/M=0.2$.}
\end{figure}

For each radius around the black hole, the azimuthal frequency
$f_\phi$ is the highest among the three characteristic frequencies
(see Fig.~1). Moreover, in general relativity, all circular orbits
with radii smaller than that of the innermost stable circular orbit
(ISCO) are unstable. This characteristic radius, at which the radial
epicyclic frequency becomes zero, depends only on the mass $M$ and
spin $a/M$ of the black hole and is given by (Bardeen et al.\ 1972)
\begin{equation} r_{\rm
ISCO}=\left\{3+Z_2\mp\left[(3-Z_1)(3+Z_1+2Z_2)\right]^{1/2}\right\}
   \frac{GM}{c^2}\;,
\end{equation}
where 
\begin{eqnarray}
Z_1 &=& 1+\left(1-\frac{a^2}{M^2}\right)^{1/3}\nonumber\\
& & \qquad
   \left[\left(1+\frac{a}{M}\right)^{1/3}+
   \left(1-\frac{a}{M}\right)^{1/3}\right]\;,\\
Z_2 &=& \left(3\frac{a^2}{M^2}+Z_1^2\right)^{1/2}\;,
\end{eqnarray}
and the sign in equation~(1) depends on whether the orbit is prograde
or retrograde with respect to the black-hole spin. Because the
azimuthal frequency decreases monotonically with radius,
\begin{equation}
f_\phi(r)=\frac{(GM)^{1/2}}{2\pi}\left[r^{3/2}\pm\frac{a}{M}
   \left(\frac{GM}{c^2}\right)^{3/2}\right]^{-1}\;,
\end{equation}
it follows that stable circular orbits can exist only with azimuthal
frequencies less than $f_\phi(r_{ISCO})$. It is expected that for any
model of quasi-periodic oscillations in accretion flows, the maximum
frequency of any {\em lowest-order, linear\/} hydrodynamic mode in the
accretion flow will also be smaller than the azimuthal Keplerian
frequency evaluated at the innermost stable circular orbit, i.e., that
\begin{equation}
f_{\rm QPO}\le f_\phi(r_{\rm ISCO})\;.
\label{eq:fmax}
\end{equation}

\begin{figure}
  \includegraphics[height=.3\textheight]{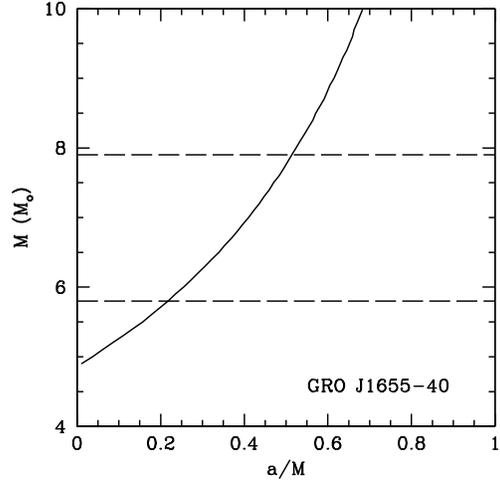} 
  \caption{{\em Dashed Lines:\/} the dynamical measurement of the
black-hole mass in the source GRO~1655$-$40. {\em Solid Line:\/} the
minimum spin parameter $a/M$ for each black-hole mass, for which the
observed 450~Hz QPO can be produced as a Keplerian frequency of a
stable orbit.  According to this argument, the black hole in
GRO~1655$-$40 is spinning with $a/M>0.2$ (after Strohmayer 2001a).}
\end{figure}

For a given observed QPO frequency, inequality~(\ref{eq:fmax})
results in a minimum spin parameter $a/M$ for the black hole as a
function of its mass $M$. This is illustrated in Fig.~2 for the black
hole in GRO~J1655$-$40 and the observed maximum QPO frequency of
450~Hz (see Strohmayer 2001a). The combination of
inequality~(\ref{eq:fmax}) (solid line) with the black-hole mass
(dashed line) measured from observations of the binary orbit (Shabaz
et al.\ 1999) provides a firm lower bound on the spin parameter for
this black hole of $a/M\ge 0.3$ (Strohmayer 2001a).

The above argument provides the most model-independent constraint on
the spin of a black hole based on its rapid variability
characteristics. Similar arguments have been used also in constraining
the masses of neutron stars showing high-frequency QPOs (Miller et
al.\ 1998) and of AGN exhibiting quasi-periodic variability (e.g.,
Iwasawa et al.\ 1998). They are general and model independent;
however, they rely on two rather restrictive assumptions: {\em (i)\/}
that we are observing the lowest-order, linear modes in the accretion
flows and {\em (ii)\/} that none of these modes occurs at a frequency
higher than the local Keplerian frequency. Even though it is probably
safe to make the second assumption, the frequency ratios of
constant-frequency QPOs in black-hole candidates (see below) cast
serious doubts on the validity of the first assumption and hence on
the generality of this argument.

\subsection{QPOs as Linear Modes}

\begin{figure}
  \includegraphics[height=.3\textheight]{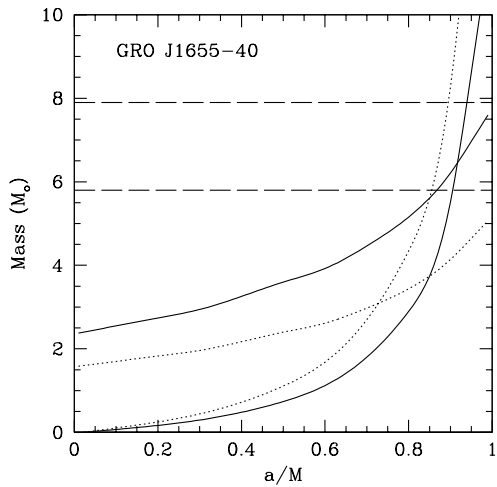} 
\caption{{\em Dashed Lines:\/} the dynamical measurement of the 
black-hole mass in the source GRO~J1655$-$40. The intersection of the
solid lines corresponds to the black-hole mass and spin for which the
observed 450~Hz and 300~Hz QPOs are identified as the lowest-order $g$
and $c$ modes, whereas this identification is reversed in drawing the
dotted lines.  According to this explanation, the black-hole spin is
$a/M\sim 0.9$ (after Wagoner et al.\ 2001).}
\end{figure}

A more accurate, albeit model dependent, measurement of the spin of a
black hole can be achieved by identifying the two observed QPOs with
the frequencies of particular linear global modes in the accretion
flows.

The excitation and trapping of global modes in hydrodynamic flows
around black holes, when the effects of magnetic fields and radiation
forces have been neglected, has been studied extensively over the last
twenty years (see Kato 2001 and references therein). In particular,
three types of modes have been identified as potential sources of
quasi-periodic variability:\newline
\noindent $\bullet$ the $g$-modes: these are inertia-gravity modes
that occur at a frequency $f_{\rm g}\simeq \kappa\pm mf_\phi$
(Perez et al.\ 1997; but see Li et al.\ 2003);\\
\noindent $\bullet$ the $p$-modes: these are inertia-acoustic modes
and are not expected to produce significant modulation of the X-ray
flux (Ortega-Rodriguez et al.\ 2002);\\
\noindent $\bullet$ the $c$-modes: these are corrugation modes
that occur at a frequency $f_{\rm c}\simeq f_\phi-f_\perp$
(Silbergleit et al.\ 2001).\newline
The frequencies of all these modes depend primarily on the mass and
spin of the black hole and weakly on the hydrodynamic properties of
the accretion flows. 

Identifying an observed QPO frequency with one of these modes leads to
a relation between the values of black-hole mass and spin for which
the observed frequency can be obtained. Identifying two observed QPO
frequencies with two different modes leads to a single pair of
black-hole mass and spin. This is illustrated in Fig.~3 for the 300~Hz
and 450~Hz QPOs observed from the black hole in GRO~J1655$-$40. The
intersection of the dotted lines corresponds to the black-hole mass
and spin for which the lowest-order $c$-mode frequency is $\simeq
300$~Hz and the lowest-order $g$-mode frequency is $\simeq 450$~Hz. On
the other hand, the intersection of the solid lines corresponds to the
black-hole mass and spin for which the identification of the observed
frequencies with the above modes has been interchanged. As long as
this model is correct, there are no free parameters in obtaining these
two pairs of values for the black-hole mass and spin!

The dynamical measurement of the mass of the black hole provides, in
principle, an independent test of the validity of this argument. In
the case of GRO~J1655$-$40 (Fig.~3), the dynamically measured mass
(area between the dashed lines) is consistent with one of the two
masses inferred from the QPO identification and the corresponding spin
parameter of the black hole is $a/M\simeq 0.9$ (Wagoner et al.\
2001). The large value of the black-hole spin may be related to the
fact that GRO~J1655$-$40 is one of the few known microquasars in the
galaxy.

\subsection{QPOs as Resonances}

Both of the arguments presented in the previous subsections rely on the
assumption that the observed QPOs are generated at the frequencies of
linear hydrodynamic modes of the accretion flows. However, in the
three systems for which pairs of constant-frequency QPOs have been
detected (including GRO~J1655$-$40), the ratios of the frequencies of
the QPOs are consistent with being equal to the ratios of small
integers (3/2 and 5/3; see Strohmayer 2001a, 2001b; Remillard et al.\
2002). This property has led to the suggestion that the observed QPOs
do not correspond to independent, linear modes but rather correspond
to pairs of non-linear modes in resonance (Abramowicz \& Kluzniak
2001; see also Abramowicz, this volume).

In contrast to the trapping of global, linear modes in the accretion
flows discussed earlier, in this interpretation, pairs of oscillatory
modes attain high amplitudes at certain radii in the accretion flows where
their frequencies have ratios equal to the ratios of small
integers. Because the radius of resonance is a free parameter, the
identification of two QPO peaks with the resonance between two
frequencies does not lead to a single pair of values for the
black-hole mass and spin but rather to a monoparametric family of
possible values (the radius of resonance being the
parameter). Moreover, additional freedom exists, at present, because
different characteristic frequencies may be at resonance in various
ways. This is illustrated in Fig.~4, where the observed 450~Hz and
300~Hz QPOs in the black hole in GRO~J1655$-$40 are identified as a
2:1, 3:1, or 3:2 resonance between the azimuthal and radial epicyclic
frequencies in the equatorial plane of the surrounding
spacetime. Clearly, only the first two of the three alternatives
result in black-hole masses consistent with the dynamical mass
measurement of this source and, for those two alternatives, the
inferred spin parameter of the black-hole is $a/M\sim 0.2-0.7$
(Abramowicz \& Kluzniak 2001; a larger spin is inferred if the
resonance is assumed to occur between the vertical and radial
epicyclic frequencies; see Abramowicz, this volume).

\begin{figure}
  \includegraphics[height=.3\textheight]{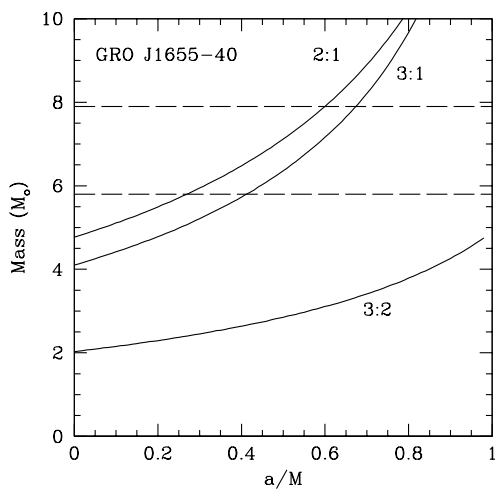}
\caption{{\em Dashed Lines:\/} the dynamical measurement of the 
black-hole mass in the source GRO~J1655$-$40. {\em Solid Lines:\/} the
combination of black-hole mass and spin for which the observed 450~Hz
and 300~Hz QPOs can be identified as a 2:1, 3:1, or 3:2 resonance
between the Keplerian and radial epicyclic frequencies. According to 
this identification, the black-hole spin is $a/M\sim 0.2-0.7$ (after
Abramowicz \& Kluzniak 2001).}
\end{figure}

\subsection{Measuring Black-Hole Spins}

The above line of arguments makes clear the potential of measuring the
spins of black holes using their rapid variability properties.
Contrary to many other complex aspects of compact objects, nature has
been kind enough to provide black holes with quasi-coherent X-ray flux
modulation at nearly constant frequencies. This constancy strongly
argues in favor of identifying the observed QPO frequencies with
dynamical frequencies that depend only on the mass and spin of the
black holes, making the inference of the latter only weakly model
dependent. However, for a precise measurement of the spin of a black
hole to be achieved, a number of important issues need to be
resolved:\newline 
$\bullet$ {\em Are linear, super-Keplerian modes possible in accretion
disks?} The least model-dependent arguments that may lead to a
measurement of the spin of a black hole rely entirely on the
assumption that such modes cannot exist.\newline
$\bullet$ {\em Are the observed QPO frequencies in ratios of small
integer numbers?}  Non-linear coupling of modes appears impossible to
neglect.\newline
$\bullet$ {\em What are the effects of magnetic fields on the
frequencies of linear modes in accretion disks?} Only a handful of
calculations of linear modes have been performed when the effects of
magnetic fields are not neglected (Gammie \& Balbus 1994; Curry \&
Pudritz 1995), even though it is thought that MHD turbulence provides
the main reason why black holes accrete at the high observed
rates.\newline
$\bullet$ {\em Why are there no QPOs in numerical simulations of MHD
accretion disks?} Current limitations in computer power allow only for
simulations of geometrically thick disks in the absence of radiation
cooling, both of which may be responsible for the lack of coherent
large-scale oscillations (see, e.g., Hawley \& Krolik 2001, 2002;
Armitage et al.\ 2001).

\section{Strong-Field Gravity}

The identification of observed QPO frequencies with dynamical
frequencies in the spacetimes of black holes provides the strongest
evidence for the existence of black holes in the universe.  Moreover,
the potential of measuring the spins of nearly-maximally rotating
black holes (as inferred, e.g., in Fig.~3) will have important
implications for our understanding not only of the formation and
evolution of black holes, but also of spin related phenomena such as
jets and outflows. All the above carry the potential of providing the
most constraining test of General Relativity in the strong-field
regime todate.

\subsection{The Strongest Case for Black Holes}

Quasi-periodic oscillations in the X-ray flux of black holes retain
phase coherence for many tens of cycles (see, e.g., Strohmayer 2001a,
2001b). Moreover, the QPOs discussed in the previous section have
properties (e.g., coherence, amplitude, photon-energy dependence,
etc.) that are reproducible and correlated with the spectral states of
the sources. These have two direct consequences: {\em
(i)\/} the regions responsible for the quasi-periodic modulations have
to be smaller than
\begin{eqnarray}
\frac{R_{\rm QPO}}{R_{\rm S}}&\le& \frac{c^3}{GMf_{\rm QPO}}
\nonumber\\
&=& 96.3
\left(\frac{f_{\rm QPO}}{300~\mbox{Hz}}\right)^{-1}
  \left(\frac{M}{7 M_\odot}\right)^{-1}\;,
\end{eqnarray}
where $R_{\rm S}$ is the Schwarzschild radius for an object of mass
$M$ and $f_{\rm QPO}$ is the observed QPO frequency; and {\em (ii)\/}
the modulating regions have to be arranged in an axisymmetric way
around the black hole or the QPO properties would not be
reproducible.  The most probable configuration is the one in which the
modulating region engulfs the central object and hence the observation
of a 300~Hz QPO from the $\simeq 7 M_\odot$ source GRO~J1655$-$40 
strongly suggests that all seven solar masses are packed within less
than one hundred Schwarzschild radii.

The above constraint can become significantly tighter if the observed
QPO frequency is required to be lower than the azimuthal frequency at
the radius where it is produced (see discussion in previous
section). This is illustrated in Figure~5 for the case of the black
hole in GRO~J1655$-$40. Identifying the observed 450~Hz QPO with a
frequency smaller than the azimuthal frequency at any radius implies
that the central object is constrained to reside within $\sim 4-6$
Schwarzschild radii. This constraint is several orders of magnitude
tighter than the one inferred from orbits of stars around the black
hole in the center of the Milky Way (e.g., Sch\"odel et al.\ (2002),
where the central object is constrained to be at most $\sim 1000$
Schwarzschild radii.

\begin{figure*}
  \includegraphics[height=.3\textheight]{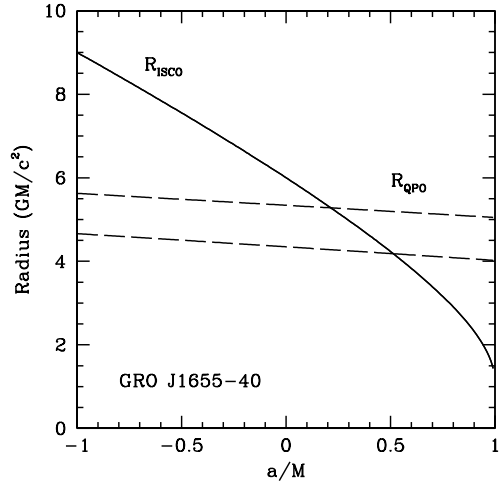} 
\caption{The dashed lines show the upper bound on the size of the
$5.8-7.9 M_\odot$ compact object in GRO~J1655$-$40, as a function of
its spin, imposed by the requirement that the 450~Hz QPO frequency is
at most the azimuthal frequency at some radius in the black-hole
equatorial plane. This is the tightest observational constraint for
the size of a black hole in the universe. The solid line shows the
radius of the innermost stable circular orbit.}
\end{figure*}

\subsection{Testing Einstein Gravity}

A definitive proof for the existence of an event horizon around a
compact object is by itself a test of the strong-field regime of
general relativity. However, a large number of alternatives to general
relativity allow also for the Schwarzschild solution and hence, in the
limit of very slow rotation, are indistinguishable from it. The
mapping of the spacetime of a {\em rotating\/} black hole using the
properties of quasi-periodic oscillations appears to be the most
promising way of testing directly the particular form of Einstein's
theory of general relativity in the strong field regime.

\begin{figure*}
  \includegraphics[angle=-90,width=15cm]{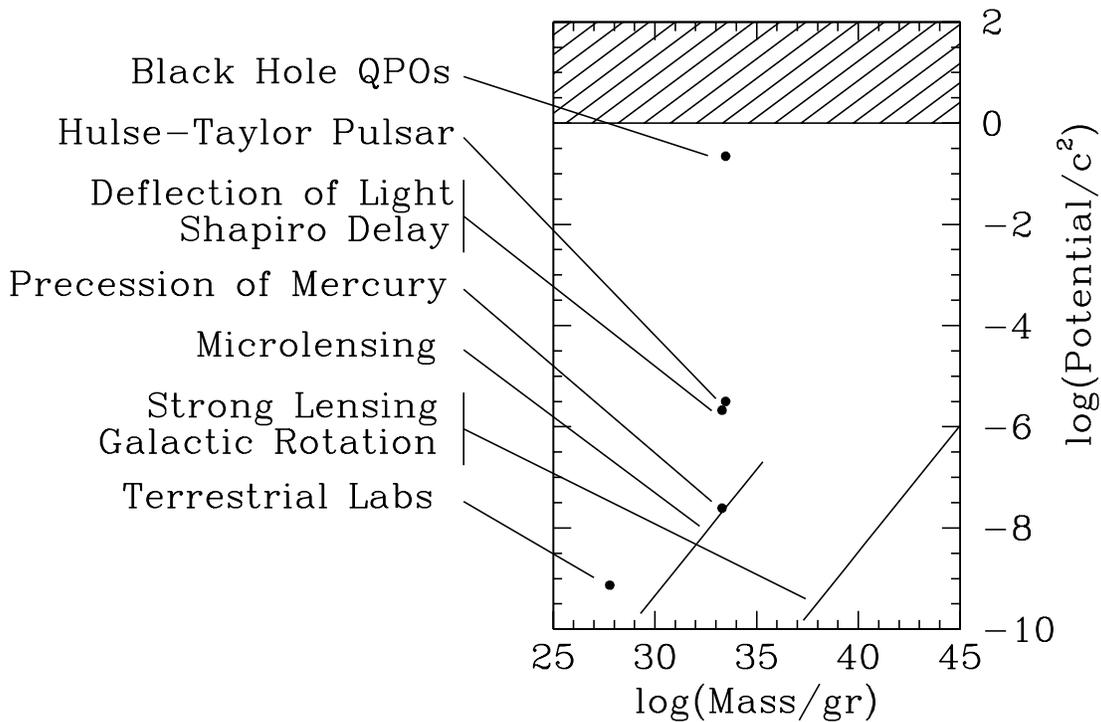} 
\caption{The gravitational potential that is probed by different
tests of gravity plotted against the mass that generates it.
Black-hole variability studies have the potential of testing gravity
theories at fields that are up to five orders of magnitude stronger
compared to any other tests.}
\end{figure*}

There is very little experimental evidence today for the behavior of
gravity in the strong-field regime (see, however, Damour \&
Esposito-Farese 1993, 1996; DeDeo \& Psaltis 2003). This fact is
illustrated in Fig.~6, where various probes and test of gravity
theories are displayed on a parameter space consisting of the mass of
the gravitating object and the potential ($\sim GM/Rc^2$) experienced
by the test particles involved in the particular tests. Clearly, all
current tests involve gravitational potentials that are at least five
orders of magnitude weaker than the potential probed by QPO
observations of black holes. These tests show that general relativity
describes the weak-field regime of our universe to within at least
$\sim 10^{-4}$ (as inferred, e.g., from tests of Brans-Dicke gravity;
see Will 2001). As a result, deviations of order unity from the
predictions of general relativity may appear in the strong-field
regime, even though the corresponding deviations may be hidden from
the weak-field tests.

Theories of gravity that are derived from an Einstein-Hilbert action
that is more general than that of general relativity are actively
being considered as explanations of a number of puzzles in cosmology,
such as the presence and magnitude of the cosmological constant (see,
e.g., Carrol 2001). Theories with profound implications for the
properties of solar-mass black holes will also affect the evolution of
the universe at the time of nucleosynthesis (cf.\ Santiago et al.\
1997). As a result, the properties of quasi-periodic black-hole
variability may provide a probe of strong-field gravity that is
complementary to cosmology. Analyses of such alternatives have been
already been performed in the case of observables from neutron-star
systems (Damour \& Esposito-Farese 1996; DeDeo \& Psaltis 2003).

\subsection{Constraining The Size of Large Extra Dimensions}

The complete theory of gravity may, of course, differ from general
relativity not only in the form of the action from which it is derived
by also in more profound ways, such as the number and scale of
spacetime dimensions in which it is defined. The predicted properties
and evolution of astrophysical black holes in such theories differs
significantly from general relativity and hence may be directly
testable.

Fig.~7 illustrates an example of constraining the size of a large
extra dimension using the properties of the black hole in
XTE~J1118$+$480 (following Emparan et al.\ 2002). In theories with one
large extra dimension, black holes evaporate significantly faster
compared to what is predicted by general relativity. As a result, the
inferred age of 240~Myr for this $\sim 6.4M_\odot$ black hole imposes
an upper limit on the size of the extra dimension of $\sim 10^{-5}$~m,
which, albeit infinite compared to the Planck length, is at least an
order of magnitude more stringent than the best upper limit obtained
with table-top experiments on the sub-mm behavior of Newton's law of
gravity.

\section{Conclusions}

Studies of black-hole variability probe the strongest field regime of
gravity that is possible for observers outside the horizon of the
black hole. They offer the potential of proving the existence of black
holes in the universe, measuring their spins, and testing gravity
theories in regimes that is unattainable by local experiments and can
complement cosmological probes.

\begin{figure}
  \includegraphics[height=.3\textheight]{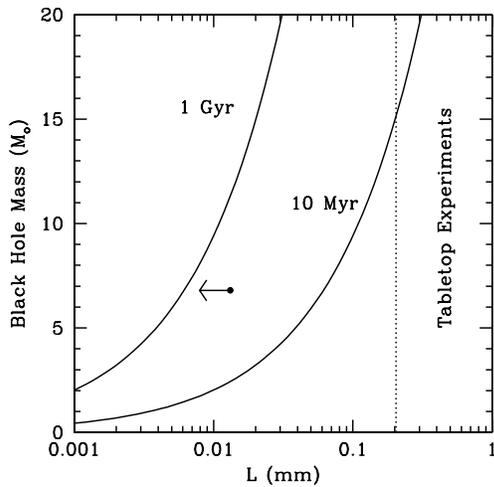} 
\caption{The solid lines are contours of constant lifetime 
for black-holes of stellar mass in theories with a large extra
dimension, as a function of the size of the extra dimension. Table-top
experiments provide a bound of $L\le 0.2$~mm. The dot represents the
upper bound on the size of the large extra dimension imposed by
inferring an age of at least 240 Myr for the $6.8\pm 0.4 M_\odot$
black hole in XTE~J1118$+$480 (after Emparan et al.\ 2002).}
\end{figure}


\begin{theacknowledgments}
It is a pleasure to thank Tomaso Belloni, Chi-Kwan Chan, Simon DeDeo,
Mike Nowak, Feryal \"Ozel, Martin Pessah, and Michiel van der Klis for
all the discussions over the last several years that have helped me
appreciate the various ways of using compact object variability in
measuring black hole spins and in testing strong-field gravity.
\end{theacknowledgments}




\end{document}